\documentclass[pre,twocolumn,showpacs,floatfix]{revtex4}

\usepackage{amsmath}
\usepackage{amssymb}
\usepackage{graphicx}

\newcommand{\vect}[1]{\mathbf{#1}}

\newcommand\varFPhi{\frac{\delta {\mathcal F}}{\delta \phi}}

\begin{document}

\title{Fracture in Mode I using a Conserved Phase--Field Model}
\date{\today}

\author{L. O. Eastgate}
\author{J. P. Sethna}
\author{M. Rauscher}
\author{T. Cretegny}
\affiliation{Laboratory of Atomic and Solid State Physics, Cornell University,
Ithaca, NY 14853}
\author{C.-S. Chen}
\author{C. R. Myers}
\affiliation{Cornell Theory Center, Cornell University, Ithaca, NY 14853}

\begin{abstract}
We present a continuum phase-field model of crack propagation. It
includes a phase-field that is proportional to the mass density and a
displacement field that is governed by linear elastic theory. Generic
macroscopic crack growth laws emerge naturally from this model.
In contrast to classical continuum fracture mechanics simulations, our model
avoids numerical front tracking. The added phase-field smoothes the sharp
interface, enabling us to use
equations of motion for the material (grounded in basic physical principles)
rather than for the interface (which often are deduced from complicated theories
or empirical observations)\/. The interface dynamics thus emerges naturally. In
this paper, we look at stationary solutions of the model, 
mode I fracture, and also
discuss numerical issues. We find that the Griffith's threshold
underestimates the critical value at which our system fractures due to long
wavelength modes excited by the fracture process.
\end{abstract}



\maketitle


\section{Introduction}
The study of fracture is usually approached using mathematical descriptions and
numerical simulations based on empirical observations. Finite element methods
are commonly used to investigate the behavior of fractured materials on a large
scale, where the crack growth laws\cite{hodgdon:1993,david} (that is, velocity and
direction of the crack for a given stress field near the tip) are introduced 
empirically.

We present a continuum description starting from basic theoretical
assumptions. We introduce a phase-field model, originally used to describe
thermodynamic phase transitions and widely used to model
solidification\cite{solidification}, and combine it with a displacement
field. In contrast to other phase-field models of 
fracture\cite{Aranson,karma:2001} and interfacial motion in the presence
of strain\cite{Muller,Kassner}, our phase
field is conserved, representing the density of the material.
Bhate {\it et al.}\cite{Brown} study a conserved order-parameter phase-field model
in the context of stress voiding in electromigration; their dynamics is
rather different from ours, since their elastic deformations are 
quasi-statically relaxed (the limit in our theory of $\eta \to 0$, see below)\/.

\begin{figure}
	\includegraphics[bb=20 60 1732 1051,
	clip=true,width=0.45\textwidth]{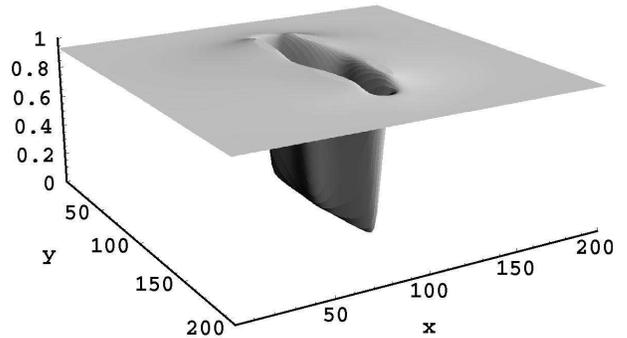}
	\caption{A three dimensional representation of the phase-field in a
	fracturing sample. The vertical axis shows the value of the phase-field
	$\phi(x,y)$, where $\phi=1$ is unstrained material and $\phi=0$ is
	vacuum. This example corresponds to the fourth contour in
	Fig.~\ref{fig:crack_contours}. The values of $x$ and $y$ are given in
	units of w/h (see App.~\ref{sec:unitless}).}
	\label{fig:3D_phiplot}
\end{figure}

The phase-field serves two main purposes. First, it smears out any sharp
interfaces, facilitating numerical convergence. Second, the model gives
equations of motion for the material rather than the boundaries, thus we avoid
dealing with a moving boundary value problem which would require numerical front
tracking. One of our main goals is to find macroscopic fracture laws. In our
model, these laws emerge naturally from the dynamics of the fields. See
Fig.~\ref{fig:3D_phiplot} for a three dimensional representation of the
phase-field in a fracturing sample.

One incentive to use a conserved-order parameter is simply that density is
conserved microscopically (apart from applications where etching or
sublimation is important). In general, a non-conserved phase-field will
give a non-zero velocity even for a straight material interface
\cite{Aranson}. This could be remedied by tuning the free energy so that the
material and vacuum have the same energy density, but then the strained region
around the crack tip would evaporate. Conserving the phase-field also gave us
insight into how to properly implement the conservation laws (see
Sec.~\ref{sec:fracture_model}).
Another option would have been to add a non-conserved field, such as damage or
dislocation density, in addition to our conserved mass. This would add
complexity without ameliorating the numerical challenges presented by the
conservation law. In future work we intend to introduce such non-conserved
state variables to model plastic flow.

The next section gives an outline of the theoretical model, presenting the main
equations. We then investigate some of the stationary solutions analytically,
and discuss their consequences. This is followed by a brief presentation of the
numerical implementation. We then measure the crack growth velocity 
as a function of external stress and explore the fracture threshold of our
model. We conclude with suggestions for future work.


\section{The Fracture Model}
\label{sec:fracture_model}
The model consists of a phase-field $\phi$ and a displacement field
$\vect{u}$. The former field is interpreted as the normalized mass density, and
typically has values between zero and one. The latter field, through its
derivatives, represents strain in the material. The model
is based on a free energy ${\mathcal F}$\/. The equations of motion locally
conserve the density $\phi$, moves it under the flow-field $\vect{u}$, and 
evolves both $\phi$ and $\vect{u}$ in the direction of the net force from 
the free energy: in particular they are constructed so that 
$d{\mathcal F}/dt<0$. The free energy is given by the integral
\begin{equation}
\label{eq:free_energy}
{\mathcal F} = \int \left(\frac{w^2}{2}|\nabla \phi|^2 + g[\phi,\epsilon]\right) dV
\end{equation}
where
\begin{equation}
\label{eq:nongrad_free_energy}
g[\phi,\epsilon]=\frac{h^2}{4}\phi^2(\phi_s[\epsilon]-\phi)^2 + \phi^2 {\mathcal
E}[\epsilon]\,.
\end{equation}
The first term in Eq.~(\ref{eq:free_energy}) is a gradient term,
energetically penalizing spatial fluctuations in the phase-field. 
The first term in Eq.~(\ref{eq:nongrad_free_energy}) is
a Ginzburg-Landau double well potential, favoring values of $\phi$ at zero and
$\phi_s[\epsilon]\equiv 1-\epsilon_{mm}$ (using the Einstein summing
convention), representing the two phases vacuum and
solid, respectively. If the material is completely unstrained the 
solid value is $\phi_s[\epsilon]\equiv 1$, otherwise this value is either 
higher (for a compressed material) or lower (for a stretched material),
where $\epsilon_{mm}$ is the the density change for small strain.
The factor $\phi_s[\epsilon]-\phi$ can be thought of as a density of
vacancies or interstitials. The parameter $h$ controls the height of the energy
barrier between the vacuum and solid phases.
The ratio of $w$ and $h$ controls the width of the solid-vacuum interface,
that is the width of the transition from $\phi=\phi_s[\epsilon]$ to $\phi=0$\/.

The next term is the elastic
strain energy density ${\mathcal E}[\epsilon]$. The elastic energy is
calculated from the strain tensor $\epsilon$, and is given by
\begin{align}
{\mathcal E}[\epsilon] &= \frac{1}{2}\sigma_{ij}\epsilon_{ij}
=\frac{1}{2}C_{ijkl}\epsilon_{kl}\epsilon_{ij}.
\end{align}
For a homogeneous, isotropic material, the tensor $C_{ijkl}$ can
be described by the two Lam\'{e} constants $\mu$ and $\lambda$ through
\begin{equation}
C_{ijkl} = \mu(\delta_{ik}\delta_{jl} + \delta_{il}\delta_{jk}) +
\lambda\delta_{ij}\delta_{kl}\,,
\end{equation}
where $\mu$ is the shear modulus and $\lambda$ is proportional to the bulk
modulus. This gives
\begin{equation}
\sigma_{ij} = \lambda \epsilon_{mm}\delta_{ij}+2\mu\epsilon_{ij} \,.
\label{eq:lame}
\end{equation}
In two dimensions, we get
\begin{equation}
{\mathcal E}[\epsilon] = \frac{\lambda}{2}\left(\epsilon_{xx} +\epsilon_{yy}\right)^2
+\mu\left(\epsilon_{xx}^2 +\epsilon_{yy}^2 +\epsilon_{xy}^2 +\epsilon_{yx}^2\right).
\label{eq:elastic}
\end{equation}
The strain field $\epsilon_{ij}$ is related to the displacement field by
\begin{equation}
\epsilon_{ij}=\frac{1}{2}\left(\frac{\partial u_i}{\partial x_j}
+\frac{\partial u_j}{\partial x_i}\right)\,.
\label{eq:epsilon}
\end{equation}
Note that in this definition of the strain field, we are ignoring geometric
nonlinearities\cite{LandauLifshitz}, which are important for large
rotations. According to Eq.~(\ref{eq:epsilon}), the divergence of the
displacement field is just the trace
of
the strain, $\nabla\cdot\vect{u}=\epsilon_{mm}$. The displacement field
$\vect{u}(\vect{x})$ is defined in the deformed or Eulerian coordinate system,
which means that $\vect{x}$ describes a location in space.  (In the undeformed
or Lagrangian description, $\vect{x}$ would describe the location of the
material before the displacement is
taken into account; Lagrangian coordinates are usually used in finite element
calculations.)
The Lam\'{e} constants are
connected through the Poisson ratio $\nu$ by $\lambda = 2\mu\nu/(1-2\nu)$, see
Appendix \ref{sec:lame}. In the case of plane strain, the addition of the
$\nabla\cdot\vect{u}$ term in the double well potential turns out to be crucial
to preserve this relation. Since the elastic energy ${\mathcal E}[\epsilon]$ is
only defined in the material (that is, where $\phi\neq 0$), the elastic term
is multiplied by a factor of $\phi^2$; thus the strain energy will go 
to zero in the vacuum.

The equations of motion we have chosen for the phase-field $\phi$ and
displacement field 
$\vect{u}$ are overdamped and Eulerian, moving the fields
along the direction of net force. The time derivative is thus proportional to
the force on the field. Physically, our model might describe fracture of a
colloidal crystal, or ``atoms in molasses''. We are therefore intermediate
between quasi-static fracture (where the crack evolution is calculated from the
static strain of the current configuration) and
dynamic fracture (with inertial effects and wave reflection at the
boundaries)\/. Specifically, our equations of motion are
\begin{subequations}
\label{eq:equations_of_motion}
\begin{equation}
\frac{\partial \phi}{\partial t} = - \nabla \cdot \vect{J}  
\quad \quad \vect{J} = - D \nabla \frac{\delta {\mathcal F}}{\delta \phi} + \phi
\frac{\partial \vect{u}}{\partial t}
\label{eq:dotphi}
\end{equation}
\begin{align}
\frac{\partial \vect{u}}{\partial t} &= -\frac{1}{\eta} \frac{{\mathbb
D}{\mathcal F}}{{\mathbb D}\vect{u}}\nonumber\\
&= - \frac{1}{\eta} \left( \frac{\delta
{\mathcal F}}{\delta \vect{u}} + \phi \nabla \frac{\delta
{\mathcal F}}{\delta \phi} \right)\,,
\label{eq:dotu}
\end{align}
\end{subequations}
where $\eta$ and $D$ are the viscosity and the diffusion constant,
respectively. Note that Eq.~(\ref{eq:dotphi}) is the continuity
equation. This means that total $\phi$, or mass, is conserved. The first term in
$\vect{J}$ is a diffusion term, while the second term makes sure that the mass
follows the motion of the displacement field. The total variational derivative
${\mathbb D}{\mathcal F}/{\mathbb D}\vect{u}$ in Eq.~(\ref{eq:dotu}) can be
found by first noting that a small change ${\Delta}\vect{u}$ in 
$\vect{u}$ results in a change in $\phi$. The new value of $\phi$ at a point
changes due to two effects: (1)~a gradient in $\phi$ dragged by a distance
${\Delta}\vect{u}$ changes it by $-(\nabla\phi) \cdot ({\Delta}\vect{u})$,
and (2)~a divergence in ${\Delta}\vect{u}$ causes a change in density
$-\phi \nabla \cdot ({\Delta}\vect{u})$. Together
these combine into the net change
${\Delta}\phi[{\Delta}\vect{u}]
	=-\nabla\cdot(\phi\, {\Delta}\vect{u})$. 
This is the continuity equation, where $\phi\, {\Delta} \vect{u}$ is the
flux of $\phi$. The total change in the free energy is then
\begin{align}
{\Delta}{\mathcal F}[{\Delta} \vect{u}] &=
\int \left(\frac{\delta{\mathcal F}}{\delta\vect{u}}
\cdot{\Delta}\vect{u}
+\frac{\delta {\mathcal F}}{\delta\phi}
	{\Delta}\phi[{\Delta}\vect{u}]\right) dV\nonumber\\
&=\int \left[\frac{\delta{\mathcal F}}{\delta\vect{u}} \cdot{\Delta}\vect{u}
-\nabla\cdot(\phi\,{\Delta}\vect{u})
	\frac{\delta {\mathcal F}}{\delta\phi} \right]
dV\nonumber\\
&=\int \left(\frac{\delta{\mathcal F}}{\delta\vect{u}} \cdot{\Delta}\vect{u}
+\phi\nabla\frac{\delta {\mathcal F}}{\delta\phi}
\cdot{\Delta}\vect{u} \right)
dV\nonumber\\
&\equiv \int \left(\frac{{\mathbb D}{\mathcal F}}{{\mathbb D}\vect{u}}
\cdot{\Delta}\vect{u}\right) dV
\end{align}
We assume that the boundary terms vanish in the integration by parts.

With the equations of motion (\ref{eq:equations_of_motion}), the
evolution of the fields are overdamped and the free energy is decreasing
in time,
\begin{align}
\label{eq:f_dot}
\frac{d}{dt}{\mathcal F} &= \int \left(\frac{\delta {\mathcal F}}{\delta
\phi}\frac{\partial \phi}{\partial t}+\frac{\delta {\mathcal F}}{\delta
\vect{u}}\frac{\partial \vect{u}}{\partial t} \right) dV \nonumber\\
&= \int \left\{\frac{\delta {\mathcal F}}{\delta \phi} \left[D \nabla^2
\frac{\delta {\mathcal F}}{\delta \phi} - \nabla\cdot\left(
\phi\frac{\partial \vect{u}}{\partial t}\right) \right]
+ \frac{\delta {\mathcal F}}{\delta \vect{u}}
\frac{\partial \vect{u}}{\partial t} \right\} dV \nonumber\\
&= \int \left[ - D\left(\nabla \frac{\delta {\mathcal F}}{\delta
\phi}\right)^2 +  \frac{\partial \vect{u}}{\partial t} \cdot \left( \phi
\nabla\frac{\delta {\mathcal F}}{\delta \phi} 
+ \frac{\delta {\mathcal F}}{\delta \vect{u}}\right) \right]
 dV \nonumber\\
&= \int \left[ - D\left(\nabla \frac{\delta {\mathcal F}}{\delta \phi}\right)^2 
	- \eta\left(\frac{\partial \vect{u}}{\partial t}\right)^2 \right]
dV \leq 0\,.
\end{align}

To solve for the fields, the functional derivatives need to be calculated
explicitly. They can be written in the convenient form
\begin{subequations}
\label{eq:variational_derivatives}
\begin{align}
\label{eq:varFU}
\frac{\delta {\mathcal F}}{\delta u_i} 
=& -\partial_j\left[\frac{\partial g}{\partial \epsilon_{ij}}\right]\\
\label{eq:varFPhi}
\varFPhi =& 
-w^2 \nabla^2\phi + \frac{\partial g}{\partial \phi}\,.
\end{align}
\end{subequations}

Finally, it should be pointed out that all the equations can be made unitless by
rescaling all the quantities involved. For the present model it is convenient to
use $w/h$ as the unit length, $h^2$ as the unit energy density, and $1/\eta$ as
the unit diffusivity. This corresponds to setting $w=1$, $h=1$ and $\eta=1$ in
Eqs.~(\ref{eq:free_energy}), (\ref{eq:dotu}), and
(\ref{eq:variational_derivatives})\/. See Appendix \ref{sec:unitless} for
information on reduced units for the quantities used in this paper.


\section{Stationary Solution}
\label{sec:stationary}
This section will calculate the profile of a stationary straight
interface between the solid and vacuum phases. Before delving into the details,
consider Eqs.~(\ref{eq:equations_of_motion})\/. A stationary solution means that
$\dot{\phi}=0$ 
and $\dot{\vect{u}}=0$. Unless we have the trivial solution where $\phi$
is constant, this implies that $\nabla[\delta{\mathcal F}/\delta\phi]=0$ and
$\delta{\mathcal F}/\delta\vect{u}=0$. Physically $\delta{\mathcal F}/\delta\phi$
can be considered a chemical potential, and a non-zero gradient will result in a
flow of material. This means that the chemical potential $p$ is a constant. Thus
to find a stationary solution, we have to require that
\begin{equation}
\label{eq:stationary_requirement}
\frac{\delta {\mathcal F}}{\delta \phi}=p \qquad \text{and} \qquad
\frac{\delta{\mathcal F}}{\delta \vect{u}}=0\,.
\end{equation}

We will find the stationary solution of a single straight interface running
perpendicular to the $x$-direction between the solid and the vacuum. We will
therefore assume 
that $\phi$ and $\epsilon_{xx}$ only vary with respect to $x$, that
$\epsilon_{yy}$ is constant, and that $\epsilon_{xy}=\epsilon_{yx}=0$. Combining
Eqs.~(\ref{eq:free_energy}) and (\ref{eq:variational_derivatives}) with the
requirement in Eq.~(\ref{eq:stationary_requirement}), we get
\begin{subequations}
\begin{align}
\label{eq:2D_phidot}
\partial_x^2\phi-\frac{\partial g}{\partial \phi} +p &= 0\\
\label{eq:2D_uxdot}
\partial_x \frac{\partial g}{\partial \epsilon_{xx}} &= 0\\
\label{eq:2D_uydot}
\partial_y \frac{\partial g}{\partial \epsilon_{yy}} &= 0
\end{align}
\end{subequations}
Integrating Eq.~(\ref{eq:2D_uxdot}) gives
\begin{equation}
\label{eq:2D_uxdot_integrated}
\frac{\partial g}{\partial\epsilon_{xx}}=C\,.
\end{equation}
For non-zero $C$, the vacuum phase ($\phi\approx 0$) can be shown to be
unstable,
so to get an interface we must set $C=0$.
Solving Eq.~(\ref{eq:2D_uxdot_integrated}) with respect to
$\epsilon_{xx}$ gives
\begin{equation}
\label{eq:2D_exx}
\epsilon_{xx} = (1-A)\left[1-\phi-(1+2\lambda)\epsilon_{yy}\right]
\end{equation}
where
\begin{equation}
A = \frac{\lambda+2\mu}{\lambda+2\mu+1/2}\,.
\end{equation}

Next we multiply Eq.~(\ref{eq:2D_phidot}) with $\partial_x\phi$. Using
Eq.~(\ref{eq:2D_uxdot_integrated}) and remembering that $\partial_x\epsilon_{yy}=0$,
we can then rewrite Eq.~(\ref{eq:2D_phidot}) as
\begin{equation}
\label{eq:2D_conserved}
\partial_x\left[\frac{1}{2}(\partial_x\phi)^2 - g[\phi,\epsilon] 
+p\phi \right] = 0 \,.
\end{equation}
Upon integration, this becomes
\begin{equation}
\label{eq:2D_conserved_integrated}
\frac{1}{2}(\partial_x\phi)^2-T[\phi] = 0\,,
\end{equation}
where
\begin{equation}
\label{eq:2D_T}
T[\phi]=g[\phi,\epsilon[\phi]]-p\phi-q\,,
\end{equation}
$q$ is the constant of
integration, and Eq.~(\ref{eq:2D_exx}) has been used to write $g[\phi,\epsilon]$ as
a function of $\phi$ only ($\epsilon_{yy}$ is considered a constant)\/.
Here $T[\phi]$ can have two minima, giving the solid and vacuum densities.

Notice that Eq.~(\ref{eq:2D_conserved_integrated}) looks like the Hamiltonian of
a classical particle at position $\phi$ and time $x$, where the first term is
the kinetic energy and $-T[\phi]$ is the potential energy.
If we want a solution that starts at a small and constant $\phi$ at
$x=-\infty$, and ends at a larger constant $\phi$ at $x=\infty$, then
$T[\phi]$ must have two stationary points with respect to $\phi$, and
these two points must have the same value. Since $T[\phi]$ is fourth
order in $\phi$, it can be written in the general form\footnote{If we had chosen
a non-zero constant of integration in Eq.~(\ref{eq:2D_uxdot_integrated}), then
Eq.~(\ref{eq:2D_exx}) would not have been linear in $\phi$, and $T[\phi]$
would not have been a simple fourth order polynomial.}
\begin{equation}
\label{eq:2D_symmetric_well}
T[\phi] = \frac{B^2}{2}(\phi-\phi_1)^2(\phi-\phi_2)^2\,.
\end{equation}
Comparing this to Eqs.~(\ref{eq:nongrad_free_energy}) and (\ref{eq:2D_T}), we
find 
that\footnote{For easy reference, this means that $\lambda+2\mu=B^2/(1-2B^2)$,
and if $\lambda=2\mu=2$ then $B=\pm 2/3$.} $B^2 = A/2$ and
\begin{equation}
\phi_{1,2}=\frac{A-\kappa\epsilon_{yy}\pm\sqrt{(A-\kappa\epsilon_{yy})^2
-\frac{2\kappa(2A-\kappa)}{1-A}\epsilon_{yy}^2}}{2A}\,,
\end{equation}
where
\begin{equation}
\kappa = A\left(1+2\lambda\right)-2\lambda\,.
\end{equation}
To second order in $\epsilon_{yy}$ this gives
\begin{subequations}
\begin{align}
\phi_1 &\approx \frac{\kappa(2A-\kappa)}{2A^2(1-A)}\epsilon_{yy}^2\\ 
\phi_2 &\approx 1-\frac{\kappa}{A}\epsilon_{yy}
-\frac{\kappa(2A-\kappa)}{2A^2(1-A)}\epsilon_{yy}^2
\end{align}
\end{subequations}
Note that $\epsilon_{yy}=0$ gives $\phi_1=0$ and $\phi_2=1$.

\begin{figure}
	\includegraphics[width=0.45\textwidth]{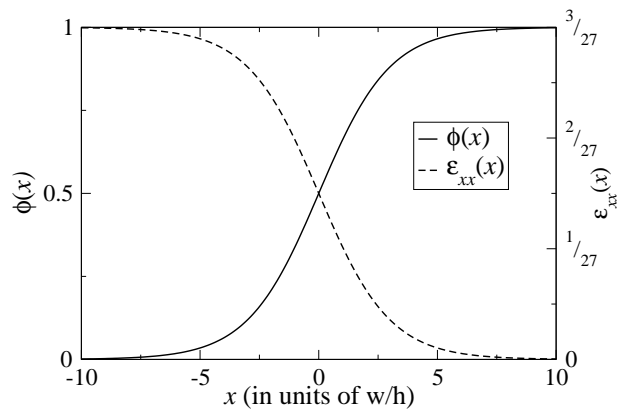}
	\caption{ A plot showing the stationary solution when
	$\epsilon_{yy}=0$, with $\lambda=2\mu=2$ and $c=0$. Note that the strain
	$\epsilon_{xx}(x)$ stays non-zero as $\phi(x)\rightarrow 0$; this is
	acceptable, since the strain does not contribute to the free energy in
	this limit (this is not true when $\epsilon_{yy}\neq 0$).}
	\label{fig:interface1D}
\end{figure}

Inserting Eq.~(\ref{eq:2D_symmetric_well}) into
Eq.~(\ref{eq:2D_conserved_integrated}) and solving for $\phi$ gives
\begin{equation}
\label{eq:2D_phi}
\phi = \delta\tanh\left[\delta B(x-c)\right] + \phi_0
\end{equation}
where
\begin{equation}
\delta = \frac{\phi_2-\phi_1}{2}\,, \qquad \phi_0=\frac{\phi_2+\phi_1}{2}\,,
\end{equation}
and the constant of integration $c$ determines the location of the
interface. In Fig.~\ref{fig:interface1D} we plot $\epsilon_{xx}$ and $\phi$ as a
function of position according to Eqs.~(\ref{eq:2D_exx}) and (\ref{eq:2D_phi})
for $\lambda=2\mu=2$, $c=0$ and $\epsilon_{yy}=0$.
Notice that the form of Eq.~(\ref{eq:2D_symmetric_well}) implies
\begin{equation}
p=2B^2 \phi_0 \phi_1 \phi_2 \qquad \text{and} \qquad q=-\frac{B^2}{2}\phi_1^2\phi_2^2\,.
\end{equation}

Consider a sample that is ${\mathcal X}$ wide and ${\mathcal Y}$ tall, with an
interface perpendicular to the $x$-direction. Let ${\mathcal X}={\mathcal
X}_1+{\mathcal X}_2$, where ${\mathcal X}_1$ is the amount of the sample that
has $\phi<\phi_0$ (vacuum), and ${\mathcal X}_2$ corresponds to $\phi>\phi_0$
(solid)\/. Then the free energy per unit length is
\begin{align}
\label{eq:2D_interface_energy}
{\mathcal F}/{\mathcal Y} &= \int_{{\mathcal X}}
\frac{1}{2}(\partial_x\phi)^2+g[\phi,\epsilon] \,dx\nonumber\\
&=\int_{{\mathcal X}} 2T[\phi]+p\phi+q\,dx\nonumber\\
&\approx\frac{4}{3}\delta^3 B +p\left(\phi_1{\mathcal X}_1+\phi_2{\mathcal
X}_2\right) +q{\mathcal X}\,.
\end{align}
The approximation is valid if the interface is far away from the boundaries of
the sample, which means that ${\mathcal X}_{1,2}\gg 1/{\delta B}$.

As a prelude to the section on numerical implementation, we should point out
some shortcomings of the current definition of the free energy. As seen above,
the limiting value of $\phi$ in the vacuum side of the interface is not zero if
$\epsilon_{yy}\neq 0$. Strictly speaking, there is no longer a vacuum, but a gas
filling the voids of the cracks. What makes this troublesome is that this
``gas'' can support shear forces. We have therefore chosen to do the numerical
simulations using $\epsilon_{yy}=0$ when measuring the fracture threshold. To
improve the model for more complex runs, the free energy could be adjusted to
assure that the value of $\phi$ in the limit of no material is zero 
for all stationary solutions. 


\section{Numerics}
\subsection{Implementation}
We have implemented Eqs.~(\ref{eq:variational_derivatives}) for a plane
strain system. Thus we can perform our simulations on a two-dimensional uniform
structured grid with periodic boundary conditions in both directions. The
periodic boundary conditions allows the use of
Fourier methods. To increase stability, we implemented a semi-implicit
scheme\cite{eyre:1998}. The linear terms can be solved analytically in Fourier
space, which increases efficiency considerably. Specifically, at each timestep
we first integrate the nonlinear terms using an explicit Euler scheme before
multiplying with the factor $\exp(-dt\nabla_{\vect{k}}^4)$, where
$\nabla_\vect{k}^2$ is the discrete version of the Laplace operator in
$\vect{k}$-space. In our case, this operator is equal to $\nabla_{\vect{k}}^2
=\sum_{i=1,2} \{[2\cos(k_i\Delta x_i)-1]/(\Delta x_i)^2\}$.
The exponential factor represents the
analytical solution to the linear part of the time derivative,
$\dot{\phi}=-\nabla^4\phi$.

\begin{figure}
	\includegraphics[width=0.45\textwidth]{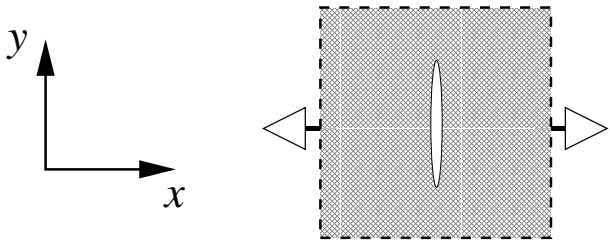}
	\caption{ A strained material with a double-ended crack. The
	hollow arrows indicate the loading direction and the dashed lines are
	the periodic boundaries.}
	\label{fig:setup}
\end{figure}

In Fig.~\ref{fig:setup} we show the setup for a double-ended crack under mode I
loading that we use in our numerical simulations.
(See Fig.~\ref{fig:3D_phiplot} for a three-dimensional representation.)

The system is initially strained in the $x$-direction (the horizontal direction
in Fig.~\ref{fig:setup}) with a uniform constant
strain $\epsilon_{xx}$. Numerically, the strain is represented through the
spatial derivatives of the displacement field $\vect{u}$, which means that there
is an inherent discontinuity in the strain field at the left and right boundary
in Fig.~\ref{fig:setup}. This
problem has been resolved using ``skew-periodic'' boundary conditions. In
essence, we identify $\vect{u}$ on the left with $\vect{u}+\Delta\vect{u}$ on
the right. 

The initial phase-field is set to the constant value that minimizes the free
energy for the uniform initial strain. A circular hole is inserted in
the middle by removing mass (that is, by tapering the phase-field to zero) in a
circular area in the center. A crack will grow if the strain
exceeds the fracture threshold. We will later compare this threshold to the
Griffith's criterion.

The fourth-order gradients in our evolution equations
\eqref{eq:equations_of_motion} make this problem numerically challenging: most
simple algorithms will go unstable at a time step which goes as the fourth
power of the grid spacing. Roughly speaking, the time step must remain smaller
than the time it takes information to pass across the footprint of the
numerical stencil (the size of the region used to calculate the gradient
terms). To make for an efficient algorithm, we paid careful attention to
minimizing this footprint area: in so doing, we found that it was important to
pay close attention to the locations of the various terms with respect to the
numerical grid. (For example, the asymmetric forward derivative of the phase
field is located at the midpoint of the
bond between two grid points). Ref.~\cite{discrete} contains further
details and suggestions. Notice that the footprint was also reduced through our
use of the semi-implicit scheme mentioned earlier, leaving only third-order
gradients to be solved numerically.

\subsection{Results}
Consider a block of material with a small flaw in it. If an increasing strain is
applied, then at a certain point the material will fail, and a crack will
form. Griffith\cite{griffith:1921} suggested that this would happen when the
strain energy released is just enough to form the two newly created crack
surfaces. Below we explore how our numerical simulations compare with Griffith's
simple model.

In most materials, the actual fracture threshold (that is, the strain energy at
which the material fractures) is higher than the Griffith's threshold. The
reason is that the strain energy is converted not only to surface energy, but
also to plastic deformation, sound emission, and heat. Our model has no plastic
deformation, but we will see that some of the strain energy is lost through long
wavelength emission similar to that of phonons in dynamic fracture.

When measuring the fracture threshold experimentally, the load or displacement
on a specimen is monotonically increased until it breaks. This is difficult to
do in our simulations, so instead we run a series of simulations, each with a
different strain as initial condition. We then measure the crack tip velocity as
a function of strain energy per unit length stored in front of the crack, and
define the fracture threshold as the energy where the velocity extrapolates to 
zero.

\begin{figure}
  \includegraphics[width=0.3\textwidth,
  bb=90 185 345 620]{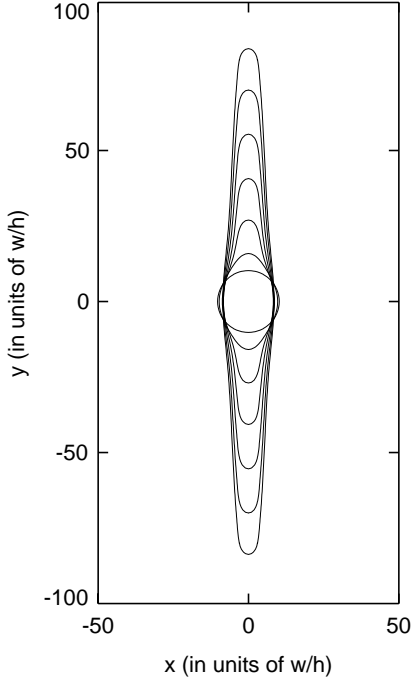}
  \caption{A section of the system showing the $\phi=0.5$ level sets at equally
  spaced time intervals. The circle in the middle is the initial
  ``crack''. After an initial slow transient the 
  crack reaches a constant velocity. When it reaches the boundary and senses its
  periodic image, the speed will increase again as the two tips coalesce and form
  a continuous crack in the $y$-direction (not shown here). This simulation
  corresponds to the data point in Fig.~\ref{fig:fracture_threshold} with
  $(\mathcal{F}/\mathcal{Y})_{\text{strain}}\approx 1$ and $v\approx 0.03$.}
  \label{fig:crack_contours}
\end{figure}

To measure the crack tip velocity we simulated a double ended crack
(see Fig.~\ref{fig:setup}) on a 
(${\mathcal X},{\mathcal Y}$)=(200,200) grid ($\Delta x=\Delta y=1$) with
periodic boundary conditions. The Lam\'{e} constants were chosen as
$\lambda=2\mu=2$, which corresponds to $\nu=1/3$. Initially, each simulation was
given a uniform
strain $\epsilon_{xx}$ in the $x$-direction and no strain $\epsilon_{yy}=0$ in
the $y$-direction. The phase-field $\phi$ was
uniformly set to the value that minimizes the free energy integrand in
Eq.~(\ref{eq:free_energy}) for a constant strain\footnote{$\phi=0$ is also a local
minimum of the free energy density, but this corresponds to the gas phase.}:
\begin{equation}
\label{eq:min_phi}
\phi = \frac{3}{4}\phi_s[\epsilon] + \frac{1}{4}\sqrt{\phi_s^2[\epsilon]
-32\,{\mathcal E[\epsilon]}}\,.
\end{equation}
To initiate the crack, a circular 
hole of radius 10 was inserted in the center of the sample; here, a ``hole''
means that we set $\phi$ to zero, being careful to make the edges smooth in
order to avoid numerical instabilities. After an initial transient period, the
crack would start to grow in the $y$-direction at a uniform rate until it
reached the boundaries. Here, the crack would sense its periodic image and speed
up as the two crack tips coalesced. Fig.~\ref{fig:crack_contours} shows part of
the grid with the $\phi=0.5$ contours as the double ended crack grows in the
$y$-direction.

\begin{figure}
  \includegraphics[width=0.45\textwidth]{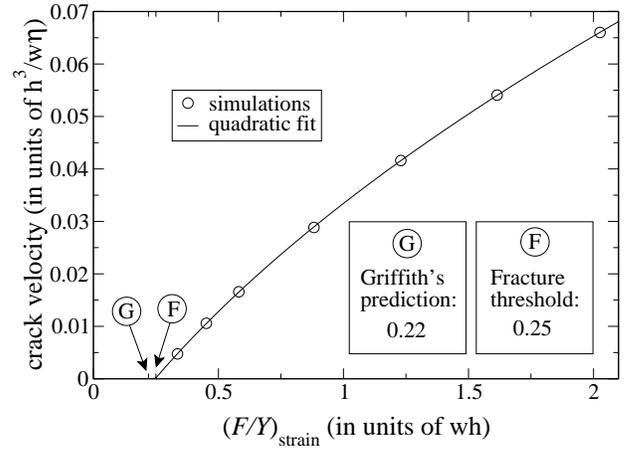}
  \caption{The graph shows the crack tip velocity against the 
  strain energy
  per unit length $({\mathcal F}/{\mathcal Y})_{\text{strain}}$ in front of the
  crack tip. The simulations were done with double ended cracks in a sample of
  size ${\mathcal X}={\mathcal Y}=200$, with $\lambda=2\mu=2$ and an initial
  strain in the $y$-direction $\epsilon_{yy}=0$. The extrapolation of the
  velocity brings it to zero at a strain energy of about $({\mathcal
  F}/{\mathcal Y})_{\text{strain}}=0.247\pm0.003$, as compared to the theoretical
  Griffith's threshold of $({\mathcal F}/{\mathcal
  Y})_{\text{surface}}=2/9\approx 0.222$.}
  \label{fig:fracture_threshold}
\end{figure}

To measure the crack tip velocity, we needed to track the crack tip. It turns
out that the free energy density has a peak in the tip area, so we decided to
define the location of this peak as the crack tip position. The velocity could
then easily be found by finding the slope of the curve describing the crack tip
position as a function of time (we used the tip growing in the positive
$y$-direction in
Fig.~\ref{fig:crack_contours}). Fig.~\ref{fig:fracture_threshold} shows a plot 
of the crack tip velocity as a function of the energy per unit length
$({\mathcal F}/{\mathcal Y})_{\text{strain}}$ in the uncracked region.
A quadratic fit of $({\mathcal F}/{\mathcal Y})_{\text{strain}}$ as a function
of the crack velocity shows that the fracture threshold is $({\mathcal
F}/{\mathcal Y})_{\text{strain}}=0.247\pm 0.003$.

A comment should be made about our use of periodic boundary conditions. Unlike
the case of dynamical fracture, our overdamped dynamics do not suffer from
elastic fields reflected from the boundaries or impinging from periodic
images. However, the periodic boundary conditions do have important effects.
As mentioned above, the velocity of the crack tip changes when the periodic
images of the crack get sufficiently close to each other. In addition, for
sufficiently thin systems (small width in the $x$-direction) and low strains
($v\approx 0$), the energy released due to the Poisson ratio is enough to
favor a crack growing in the horizontal direction, parallel to the $x$-axis
(limiting how thin we can make our rectangular region). We have minimized the
effects due to the periodic boundary conditions by using a wide system and by
only measuring the crack velocity when the tip is far from the boundaries,
though further measures might be needed for more sophisticated simulations.
Note that periodic boundary conditions make it hard to apply stresses to the
system; all the simulations presented in this paper were driven by applying an
initial strain (equivalent to applying a displacement).

Eq.~(\ref{eq:2D_interface_energy}) gives the analytical surface energy of a
stationary interface. With $\epsilon_{yy}=0$, $\lambda=2\mu=2$, 
the total free energy per length far behind the crack tip
due to the two interfaces
is\footnote{After relaxing the interface numerically, the actual value of the
surface energy is $({\mathcal F}/{\mathcal
Y})_{\text{surface}}=0.219\pm 10^{-5}$. The
reason it is not $2/9$ is that the numerical solution is slightly different from
the analytical one due to the discrete nature of the simulation.}
$({\mathcal F}/{\mathcal Y})_{\text{surface}}=2/9\approx 0.222$. If our model
were to obey the one proposed by Griffith, the crack tip velocity should tend
to zero as $({\mathcal F}/{\mathcal Y})_{\text{strain}}$ approaches $({\mathcal
F}/{\mathcal Y})_{\text{surface}}$. Apparently, there is a disparity between the
numerical fracture threshold and the analytical Griffith's threshold for our
model. Most of the energy in front of the crack tip is transferred to the newly
created crack surfaces, but some extra energy is needed to drive the crack
forward.

Upon investigation, we find a large residual strain $\epsilon_{yy}$ 
building up across the width of the sample near the height of the crack tip
during fracture.
In front of the crack, the material has contracted
in the $y$-direction, 
presumably
because of the Poisson ratio and the positive
strain perpendicular to the crack. Behind the crack, the material has been
stretched in the $y$-direction; this is necessary to make up for the compressed
material in front of the crack. 

It turns out that the difference in energy between the fracture threshold and
the Griffith's threshold roughly matches the 
residual energy stored in this deformation.
Typically, viscous effects vanish as the tip velocity goes to
zero, but we claim 
that this one goes to a constant because the width of the
deformation diverges in the limit of a stationary crack tip. Below we will give
a crude analytical argument to show that this is not a finite size 
or finite velocity effect, and
give a rough estimate of the missing energy.

Consider an infinitely long system in the $y$-direction, but with a fixed width
${\mathcal X}$. To make the system one dimensional, we will ignore any
variations in $x$. We introduce here a frame that is stationary with respect to
the crack tip, $\tilde{y}=y-vt$ and $\tilde{t}=t$, which means that $\partial_t
\rightarrow -v\partial_{\tilde{y}}$ since any derivatives with respect to
$\tilde{t}$ disappear for a stationary solution.

Both far in front of and far behind the
tip we assume that $\epsilon_{yy}=0$, which means that the displacement field
$u_y(\tilde{y})$ is constant in these areas. In the deformed region in between
(that is, where the crack tip is), the displacement 
field changes by a value $\Delta u_y=u_y(-\infty)-u_y(\infty)$ (notice that
the displacement field has a lower value in front of the tip)\/. According to
Eq.~(\ref{eq:lame}), the strain far ahead of the tip exerts a stress
\begin{equation}
\label{eq:force_in_front}
\sigma_{yy}=\lambda\epsilon_{xx}
\end{equation}
on the deformed
region.
Since there is no strain far behind the tip,
this stress must be countered by the viscous force due to the movement of the
displacement field:
\begin{align}
\label{eq:delta_u}
\lambda\epsilon_{xx}&=\int \frac{\partial u_y}{\partial t}
\,d\tilde{y}\nonumber\\
&= -v \int \frac{d u_y}{d\tilde{y}} \,d\tilde{y}\nonumber\\
&= -v\Delta u_y\,.
\end{align}

Knowing how much $u_y$ changes, we want to find its shape in the deformed
region. Using that $\partial_{\tilde{t}}u_y=0$ and $\partial_t
u_y=-\delta{\mathcal F}/\delta u_y=(\lambda+2\mu)\partial_y^2u_y$, we have that
\begin{align}
\frac{\partial}{\partial\tilde{t}}u_y 
&=\frac{\partial u_y}{\partial t} +v\frac{\partial u_y}{\partial y}\nonumber\\
&=(\lambda+2\mu)\frac{\partial^2u_y}{\partial y^2} +v\frac{\partial
u_y}{\partial y} = 0
\end{align}
A solution to this second order differential equation apart from the crack
tip position $y=v\,t$ is
\begin{equation}
\label{eq:uy_diss}
u_y(y,t) =
\begin{cases}
\Delta u_y\,, & y<vt\\
\Delta u_y \exp\left[-\frac{v}{\lambda+2\mu}(y-vt)\right]\,, & y>vt
\end{cases}
\end{equation}
where we have used that $u_y(\infty)=0$.

We can now find the energy dissipated due to the material deformation around the
crack tip. Ignoring the phase-field in Eq.~(\ref{eq:f_dot}), our crude
estimate gives the energy
dissipated per unit length as the inverse velocity times the energy dissipated
per unit time:
\begin{align}
\label{eq:energy_diss}
\left(\frac{\mathcal F}{\mathcal Y}\right)_{\text{diss}}
&=-\frac{1}{v}\frac{d}{dt}{\mathcal F}\nonumber\\
&\approx\frac{{\mathcal X}}{v}\int_0^\infty \left(\frac{\partial u_y}{\partial t}\right)^2
\,d\tilde{y}\nonumber\\
&=\frac{{\mathcal X}}{v}\int_0^\infty\left[\frac{v^2 \Delta u_y} {\lambda+2\mu}
e^{-\left(\frac{v}{\lambda+2\mu}\right)\tilde{y}}\right]^2
\,d\tilde{y}\nonumber\\
&=\frac{(\lambda \epsilon_{xx})^2 {\mathcal X}}{(\lambda+2\mu)2}\,,
\end{align}
where we used Eq.~(\ref{eq:delta_u}) in the last step. 
(To quadratic order in $\epsilon_{xx}$, this equals 
$\lambda^2/4\mu(\lambda+2\mu)$ times 
$({\mathcal F}/{\mathcal Y})_{\text{strain}}$, so the dissipation is
a fixed fraction of the total available strain energy. For our system,
the fraction is 1/8.) We thus conclude that the energy 
dissipated due to the deformation around the crack tip only depends on the
initial strain and the width of the system. In our system of width ${\mathcal
X}=200$, the strain 
corresponding to the fracture threshold is about $\epsilon_{xx}=0.026$. For
numerical comparison, we insert the strain and the other values that we used in
our simulations into Eq.~(\ref{eq:energy_diss}), and find that ${\mathcal
F}/{\mathcal Y}_{\text{diss}}\approx 0.07$, which is about twice the difference
between ${\mathcal F}/{\mathcal Y}_{\text{strain}}$ and ${\mathcal F}/{\mathcal
Y}_{\text{surface}}$. Even though we ignored both the phase-field $\phi$
and all variations in $x$, our crude estimate shows that as the velocity
goes to zero a fixed fraction of the energy in front of the crack is
used to strain the material in a diverging region around the crack tip
before it is dissipated.

We also did a simulation to verify that Eq.~(\ref{eq:uy_diss}) is of the
right form. With $({\mathcal X},{\mathcal Y})=(100,1200)$ and
$\epsilon_{xx}=0.08$ (which gives $({\mathcal F}/{\mathcal
Y})_{\text{strain}}=1.01$), we did see an exponential decay of $u_y$ in front of
the crack tip. The numerical values were slightly off, with the exponential
decay being twice as fast as predicted. This was to be expected since we have
ignored the phase-field completely, which clearly couples to the displacement
field at long distances. Our simulation area still was not long enough for
$\Delta u_y$ to saturate, but it reached about $2/3$ of the value predicted in
Eq.~(\ref{eq:delta_u})\/.

In Fig.~\ref{fig:fracture_threshold}, we plot the velocity as a function of
strain energy available for fracture $({\mathcal F}/{\mathcal
Y})_{\text{strain}}$. In the fracture literature, the mode I fracture threshold
is usually quoted in terms of the energy release rate\footnote{For plane strain,
the fracture threshold is given in terms of the stress intensity factor $K_I$, which
is related to the energy release rate by ${\mathcal G} = K_I^2 (1-\nu)/2\mu$.}:
the strain energy ${\mathcal G}$ flowing into the crack tip. This can be
measured, for example, by the use of $J$-integrals\cite{rice:1968}, see Appendix
\ref{sec:j_integral}.

From the discussion above, we conclude
that the disparity between the Griffith's prediction and the fracture
threshold as measured by $({\mathcal F}/{\mathcal Y})_{\text{strain}}$ 
is due to energy dissipation far from the crack. Thus we cannot expect
${\mathcal G}$ to equal $({\mathcal F}/{\mathcal Y})_{\text{strain}}$. Indeed,
the fracture threshold measured using a $J$-integral close to the crack 
tip should agree with Griffith's prediction.


\section{Future Work}

Fracture often occurs in crystalline materials, and it therefore seems
reasonable to add anisotropy to quantities such as mobility, surface tension, and
elastic strain energy. There might also be unwanted anisotropy included due to
the underlying grid in the numerical solver, and one might be able to repress
this effect by adding ``counter''-anisotropy in the quantities mentioned
above. In addition to anisotropy, one could add some noise; either
spatially in the Hamiltonian or initial conditions, which would break the
symmetry and add heterogeneities, or temporally, which would mimic the
effects of fluctuations due to a finite temperature.

Because of the fourth-order gradients, our simulation is rather numerically
intensive.  Using a multiresolution grid would speed up the
calculation\cite{arias:1999}. The main idea is that additional information
can be added between existing grid points where a linear interpolation would give
poor results (that is, add grid points locally where the solution has high
frequency components)\/. This way one can achieve a solution with uniform
accuracy and avoid unnecessary use of computational resources. A similar
approach can be found in\cite{provatas:1998}, where an adaptive mesh in
a phase-field solidification problem is used to add detail only at the
boundaries where it is needed.

Currently, the numerical integration of the phase-field equations is done partly
using Fourier transforms. This allows us to do all the linear parts of the
equations implicitly without solving huge linear systems. On the other hand, the
current implementations puts several limitations on the kind of simulations
that can be done; especially, it is hard to do simulations with other than
periodic boundary conditions. It thus seems beneficial to use real-space
implicit methods (and perhaps add in multi-resolution capabilities at the same
time)\/.

This paper is only concerned with
mode I fracture in two dimensions. It would seem
reasonable to try to extend the model to include mode II and mode III as
well. Mode II could be done either by shearing the model with a crack running
vertical or horizontal, or by squeezing one way and extending the other with a
strain $s$ and 45 degree crack, where $s$ is equal to the shear strain divided
by $\sqrt{2}$. One way to start exploring mode II could be to use finite
element calculations to get an initial sheared configuration, and then use the
phase-field model to relax this system. Mode III has recently been explored in
Ref.~\cite{karma:2001}, where a different, non-conserved
phase-field model is used. Ultimately, the goal is to do three dimensional
simulations incorporating all three modes of fracture, given the success of the
two-dimensional models.

 
\acknowledgments
This project was primarily supported by NSF DMR-9873214 and Norges
forskingsr{\aa}d (the Norwegian Research Council), with additional
support from the Cornell Center for Materials Research
(CCMR, a Materials Research Science and Engineering Center of the
National Science Foundation, DMR-0079992), the Cornell Theory
Center (which receives funding from Cornell University, New York
State, federal agencies, and corporate partners, as well as research
infrastructure support through NSF 9972853), and the Technology
for Education project funded by Intel. We would also like to thank
Tony Ingraffea, Paul Dawson, and Nick Bailey for helpful comments
and discussions.

\hrulefill

\appendix

\section{Reduced Units}
\label{sec:unitless}
Most of the equations in this paper are written in unitless form, where the
basic units are:
\begin{subequations}
\begin{align}
l &= w/h     && \text{(length)}\\
f &= h^2     && \text{(energy density)}\\
d &= 1/\eta  && \text{(diffusivity)}
\end{align}
\end{subequations}
Here $w^2$ is the cost of gradients of $\phi$ in Eq.~(\ref{eq:free_energy}),
$h^2/64$ is the height of the energy barrier between the 
phases in Eq.~(\ref{eq:nongrad_free_energy}), and $\eta$ is the
viscosity controlling the response of the displacement field in
Eq.~(\ref{eq:dotu})\/. The equations were made unitless by replacing all  
quantities (say, 
$x$ and $t$) by their unitless counterparts multiplied by the appropriate
combination of basic units (as in $x^\ast l$ and $t^\ast l^2/fd$, $\ast=$
unitless), and then choosing the basic units $l$, $f$ and $d$ conveniently. To
transform the quantities back from unitless form, just multiply them with their
basic unit ($x=x^\ast l$)\/.

As a reminder, the variational derivatives of the free energy and the Lam\'{e}
constants $\lambda$ and $\mu$ all have units of energy density ($f$), and the
diffusion constant $D$ has units of diffusivity ($d$)\/.

\section{Lam\'{e} Constants}
\label{sec:lame}

We have to check what the relation between $\lambda$, $\mu$ and $\nu$ is in
our model: the addition of $\phi$ changes the effective elastic constants
on long wavelengths from the values input to the free energy in Eq.~(\ref{eq:lame})\/.
In the following, two and three dimensional refer to the mathematical
dimensionality of the system, not special cases like plain strain. Start with a
rectangular sheet (2D) or block (3D) of material, with height $h$ in the
$y$-direction, and width $w$ in the $x$-direction (and in the $z$-direction if
it is in 3D)\/. The idea is then to strain the material an infinitesimal amount in
the $y$-direction until the height becomes $h^\prime$, and minimize the free energy
with respect to the new width $w^\prime$. The Poisson ratio $\nu$ is then given by
\begin{equation}
\nu = \lim_{h^\prime \rightarrow h}\left(-\frac{\epsilon_{xx}}{\epsilon_{yy}}\right)
\label{eq:nu}
\end{equation}

Before doing the model proper, we start off with a free energy that \emph{only}
includes the elastic energy:
\begin{subequations}
\begin{align}
{\mathcal F}_0 =& \int_V dV\,{\mathcal E}[\epsilon] \,.
\label{eq:fe_elastic}
\end{align}
In the scenario that is described above there is no shear, so $\epsilon_{ij}=0$ if
$i\neq j$. Further, $\epsilon_{yy}=(h^\prime-h)/h^\prime$, and
$\epsilon_{xx}=\epsilon_{zz}=(w^\prime-w)/w^\prime$ (in 2D, $\epsilon_{zz}$ is
non-existent)\/. The integrand is constant everywhere, so the integral turns into
a factor $h^\prime (w^\prime)^2$ (or $h^\prime w^\prime$ in 2D)\/. This is
inserted into Eq.~(\ref{eq:fe_elastic})\/. Next, one finds the minimum of
${\mathcal F}_0$ with respect to $w^\prime$ (keeping all the other variables
constant), and inserts into Eq.~(\ref{eq:nu}), taking the limit. Solving
for $\lambda$ does indeed give the standard answers,
for both two and three dimensions.

The next two energies that were tried, were
\begin{align}
\label{eq:F1}
{\mathcal F}_1 =& \int_V dV \frac{1}{4}\phi^2(\phi-1)^2
+\phi^2{\mathcal E}[\epsilon]\,,\\
\label{eq:F2}
{\mathcal F}_2 =& \int_V dV
\frac{1}{4}\phi^2(\phi-\phi_s[\epsilon])^2
+\phi^2{\mathcal E}[\epsilon]\,.
\end{align}
\end{subequations}
The gradient term is zero, since $\phi$ is uniform. Having introduced $\phi$,
we need to add another restriction; conservation of mass is given by $\phi_0 h
w^2 = \phi^\prime h^\prime (w^\prime)^2$ (or $\phi_0 h w = \phi^\prime h^\prime
w^\prime$ in 2D)\/. Here, $\phi_0$ is the value when the material is relaxed,
while $\phi^\prime$ is the value after the material has been stretched [it is
the latter which will be inserted into Eqs.~(\ref{eq:F1}) and
(\ref{eq:F2})]\/. We assume that $\phi_0\equiv 1$. We checked two cases for
$\phi^\prime$; one where $\phi^\prime\equiv 1$ (the density was not allowed to
change), and one where $\phi^\prime = hw/h^\prime w^\prime$ (the latter is the
2D expression; having $\phi$ change in three dimensions was quite hard to
calculate)\/.

The results after minimizing the free energy, finding $\nu$, taking the limit
and solving for $\lambda$ can be found in Table~\ref{tab:lambda}.
As one can see, adding the phase-field $\phi$ to the model also requires that
the double-well potential includes the $\epsilon_{mm}$ term (through
$\phi_s[\epsilon]$) in order to get the right relation between the material
constants.

A final comment: The model described in this paper is two-dimensional, so
the maximum range for the Poisson
ratio in two dimensions is $0<\nu<1$. This is
\emph{equivalent} to a three dimensional model under plain strain,
where the corresponding range for the Poisson ratio is 
$0<\nu<1/2$. Thus the latter expression (for three dimensional
plain strain) is used throughout the paper, often using the value $\nu=1/3$.

\begin{table}
\begin{center}
\begin{tabular}{ccccc}
Dimension & $\phi^\prime$ & ${\mathcal F}_0$ & ${\mathcal F}_1$ &
${\mathcal F}_2$\\
\hline
&&&&\\[-0.1in]
2 & $\frac{hw}{h^\prime w^\prime}$ & --- & $\lambda =
\frac{2\mu\nu}{1-\nu}-\frac{1}{2}$ & $\lambda = \frac{2\mu\nu}{1-\nu}$\\[0.1in]
2 & 1 & $\lambda = \frac{2\mu\nu}{1-\nu}$ & $\lambda = \frac{2\mu\nu}{1-\nu}$ &
$\lambda = \frac{2\mu\nu+1}{1-\nu}-\frac{1}{2}$\\[0.1in]
3 & 1 & $\lambda = \frac{2\mu\nu}{1-2\nu}$ & $\lambda = \frac{2\mu\nu}{1-2\nu}$
& $\lambda = \frac{2\mu\nu}{1-2\nu}$\\[0.1in]
\end{tabular}
\caption{ Relations between $\lambda$, $\mu$ and $\nu$ when a material
is stretched infinitesimally.}
\label{tab:lambda}
\end{center}
\end{table}

\section{The $J$-Integral}
\label{sec:j_integral}
The energy release rate ${\mathcal G}$ can be calculated for our problem
(where the crack is parallel to the $y$-axis) using the $J_y$ component of the
$J$-integral. Instead of performing the line-integral, it is common to convert
it to an area-integral for increased accuracy when doing the integral
numerically. The area integral is defined as
\begin{equation}
J_y = -\int_A \Omega(x,y) dx dy
\end{equation}
where
\begin{equation}
\Omega(x,y) =  {\mathcal E}[\epsilon]\frac{\partial q}{\partial y} -
\sigma_{ij}\frac{\partial u_i}{\partial y}\frac{\partial q}{\partial x_j}
\end{equation}
Here $q$ is a function that is unity around the crack tip and zero
outside. Notice that if q is constant in a region, $\Omega(x,y)\equiv 0$, so in
effect the line integral is replaced by a ``thick line'' integral, where the
``thick line'' exists everywhere q has a gradient.

A small complication is that our $\Omega(x,y)$ is given in deformed coordinates
$\xi_x(x,y)\equiv x+u_x(x,y)$ and $\xi_y(x,y)\equiv y+u_y(x,y)$:
\begin{equation}
\label{eq:j_integral_deformed}
J_y = -\int_{{\tilde A}}\tilde{\Omega}(\xi_x,\xi_y)
\left|{\mathcal J}(\xi_x,\xi_y)\right| d\xi_x d\xi_y
\end{equation}
where ${\mathcal J}$ is the Jacobian given by
\begin{equation}
{\mathcal J}(\xi_x,\xi_y) = \left[
\begin{array}{cc}
\frac{\partial x}{\partial \xi_x} & \frac{\partial x}{\partial \xi_y}\\
\frac{\partial y}{\partial \xi_x} & \frac{\partial y}{\partial \xi_y}
\end{array}\right]
\end{equation}
Using the identity ${\mathcal J}(\xi_x,\xi_y)\cdot{\mathcal J}(x,y) = I$, one gets
\begin{align}
|{\mathcal J}(\xi_x,\xi_y)| =& \frac{1}{|{\mathcal J}(x,y)|}\nonumber\\
=& \frac{1}{\partial_x\xi_x\partial_y\xi_y-\partial_y\xi_x\partial_x\xi_y}\nonumber\\
=& \frac{1}{(1+\epsilon_{xx})(1+\epsilon_{yy})-\partial_y u_x\partial_x u_y} \,.
\end{align}
The energy release rate can thus easily be computed using
Eq.~(\ref{eq:j_integral_deformed}), where $\xi_x$ and $\xi_y$ are the usual
coordinates on 
the phase-field grid.

\end{document}